\documentclass[twocolumn,amsmath,aps]{revtex4}
\usepackage{graphicx}

\newcommand{\nl}{\nonumber \\}

\newcommand{\cf}{cf.\ }

\newcommand{\be}{\begin{equation}}
\newcommand{\ee}{\end{equation}}
\newcommand{\bea}{\begin{eqnarray}}
\newcommand{\eea}{\end{eqnarray}}

\newcommand{\Eq}[1]{Eq.\,(\ref{#1})}
\newcommand{\Eqs}[1]{Eqs.\,(\ref{#1})}

\newcommand{\la}{\langle}
\newcommand{\ra}{\rangle}
\newcommand{\dg}{\dagger}

\newcommand{\ti}{\tilde}
\newcommand{\mb}{\mbox}

\begin{document}
\bibliographystyle{aip}


\title{ Quantum measurement of a solid-state qubit:
        A unified quantum master equation approach revisited }

\author{Xin-Qi Li$^{1,3}$, Wen-Kai Zhang$^{2}$,
    Ping Cui$^3$, Jiushu Shao$^2$,
        Zhongshui Ma$^4$, and YiJing Yan$^3$}

\affiliation{$^1$National Laboratory for Superlattices and Microstructures,
         Institute of Semiconductors,
         Chinese Academy of Sciences, P.O.~Box 912, Beijing 100083, China \\
        $^2$State Key Laboratory of Molecular Reaction Dynamics,
     Institute of Chemistry,
         Chinese Academy of Sciences, Beijing 100080, China \\
        $^3$Department of Chemistry, Hong Kong University of
    Science and Technology,
         Kowloon, Hong Kong \\
        $^4$Department of Physics,
      Peking University, Beijing 100871, China }
\date{\today}

\begin{abstract}
    Quantum measurement of a solid-state qubit by a mesoscopic detector
is of fundamental interest in quantum physics and an essential issue
in quantum computing.
In this work, by employing a unified
quantum master equation approach constructed in
our recent publications, we study the
measurement-induced relaxation and dephasing
of the coupled-quantum-dot states measured by a quantum-point-contact.
Our treatment pays particular attention on the detailed-balance relation,
which is a consequence of properly accounting for the energy exchange
between the qubit and detector during the measurement process.
As a result, our theory is applicable to measurement at arbitrary voltage
and temperature.
Both numerical and analytical results for the qubit relaxation and
dephasing are carried out, and new features are highlighted
in concern with their possible relevance to future experiments.
\end{abstract}
\vspace{2ex}


\maketitle

\section{Introduction}
\label{thsec1}

Measuring a two-state quantum system
(qubit) typically represents the long-standing
and still controversial issue in quantum measurement.
Ideally, based on the standard Copenhagen postulate,
quantum measurement is described as a {\it wavefunction collapse},
i.e., projects the qubit state to one of the possible eigenstates of
the observed quantity with state-dependent probabilities.
However, in practice, any realistic measurement is
performed by a realistic device
that itself is a physical system.
Response of the measured system to the
measuring device is in general a non-trivial problem,
which has attracted considerable attention in recent years [1-18].
This renewed interest also stems from the
rapidly developing field of quantum
computing, since the quantum measurement procedure is
needed, for instance, at the end of computation to read out the
final results, or even in the course of computation for the
purpose of error correction.

A possible implementation of the two-state quantum measurement
is to consider a charge qubit being measured by a charge sensible detector,
such that the transport current in the detector
carries information of the measured qubit.
The charge qubit can be either an extra electron
stored in coupled quatum dots (CQDs) \cite{Li01},
or an extra cooper pair in superconducting box \cite{Sch98,Nak99,Sch02},
meanwhile the detector can be a quantum-point-contact (QPC) [1-6,12,16],
or a single-electron-transistor (SET) [7-11,17].
In these studies, in addition to theoretical discussions [1-10,16,17],
experimental results have also been reported [11-13].

To study the effects of measurement on a quantum system,
the standard procedure is to trace out
the microscopic degrees of freedom of the
detector, which would result in a reduced
description in terms of quantum master
equation (QME) for the relaxation and dephasing of the measured system.
In the seminal work by Gurvitz \cite{Gur97}, the quantum
measurement of the charge state in coupled QDs by a QPC, was
studied based on a reducing procedure from the many-particle
wavefunction of the entire qubit-plus-detector system.
This approach was also applied to study the breakdown of the
Anderson localization in the presence of
quantum measurement \cite{Gur00},
and its conditional version was exploited
to analyze the readout of the detector \cite{Moz02,Gur03}.
By an alternative means,
Goan {\it et al} derived a Lindblad QME for the same measurement setup,
based on which a quantum trajectory description was developed for the single
continuous measurement \cite{Goa01}.
The Lindblad QME obtained by
Goan {\it et al} has also been demonstrated to be
equivalent to the Bloch equations derived by Gurvitz \cite{Goa01}.
However, we notice that their master equations would
inevitably lead to certain  peculiar features
such as the {\it always} equal occupation probabilities
on  individual dots (sites) in the asymmetric-qubit (disorder-chain)
after the completion of dephasing and relaxation \cite{Gur97,Gur00}.

In this work, we revisit this well-defined quantum measurement problem
by employing a unified Markovian QME approach \cite{Yan98,Li02,Yan00}.
We pay particular attention
on the detailed-balance relation, which properly accounts for
the energy exchange
between the qubit and detector during the measurement process.
Consequently, our approach
is valid at arbitrary measurement voltage and temperature.
It will show that
the results in Refs.\ \onlinecite{Gur97}
and \onlinecite{Goa01} break down at small voltage
and the peculiar features in
Refs.\ \onlinecite{Gur97} and \onlinecite{Gur00}
survive only in high voltage regime.
Note that the measurement voltage can
in a certain sense be interpreted as
an effective temperature \cite{Moz02};
thus the equal occupation probabilities
on individual states may be viewed as
the result of an effectively strong thermalization.
To our knowledge, this kind of clarification lacks so far in literature.
Recently, similar quantum measurement under arbitrary voltage
is analyzed in terms of the noise spectrum
of the detector output signal \cite{Sto02,Shn02,Sch02}.
In the large voltage regime, the
noise spectrum is symmetric; and in small voltage regime, the noise spectrum
becomes asymmetric. This change of spectral shape indicates a transition
from classical to quantum. In the quantum regime,
it is right the {\it energy exchange}
between the qubit and detector that leads to
the asymmetry of noise spectrum.
Our present work, which deals with
the measurement-induced qubit dephasing
and relaxation under arbitrary voltage and temperature,
thus provides an alternative perspective
to elucidate  the nature of energy exchange and its
importance  in describing quantum measurement.

 The remainder of this paper is organized as follows.
In the context of the considered quantum measurement model,
we outline in Sec.\ II the derivation
of the QME for the reduced dynamics of the qubit.
The detailed-balance property of our QME
and its relation to the energy exchange processes
accompanying the quantum measurement
will be elaborated in the Appendix.
In Sec.\ III,
numerical results of the relaxation
and dephasing behaviors of the measured qubit
are presented, and discussions are highlighted to
some new features resulting from the detailed balance.
Section IV contributes to the relaxation and dephasing rates where
the newly derived analytical formulas clearly describe effects of various
measurement parameters, such as the bias voltage
across the QPC and the temperature in the electronic reservoirs.
Finally, in Sec.\ V we summarize the main results
and implications of this work.

\begin{figure}\label{Fig1}
\begin{center}
\centerline{\includegraphics [scale=0.5] {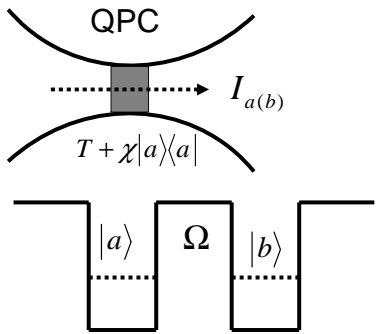}} \caption{
Schematic diagram of a solid-state qubit (coupled quantum dots)
being measured by a quantum point contact (QPC). Occupation of the
extra electron in different dots would have distinct influence on
the transport current through the QPC, that makes it possible to
draw out the qubit state information. On the other hand, as a
result of backaction of the detector, quantum coherence of the
qubit state would be destroyed. }
\end{center}
\end{figure}

\section{QME approach for quantum measurement: formal result}
\label{thsec2}
\subsection{Model description}
\label{thsec2A}

Following the previous work \cite{Gur97,Goa01,Sto99}, we
consider here a coupled quantum dots (i.e.\ a solid-state qubit) measured
by a quantum point contact, as schematically shown in Fig.\ 1.
To present a microscopic description for the measurement,
assume the Hamiltonian of the entire qubit-plus-reserviors system as
\begin{subequations} \label{H1}
\be \label{H1a}
 H   =  H_{\rm qu} + H_{\rm res} + H'  ,
\ee
with
\bea
 &&H_{\rm qu} =  \epsilon_a |a\ra\la a| + \epsilon_b |b\ra\la b|
             + \Omega(|b\ra\la a|+|a\ra\la b|) ,
\label{H1b} \\
 &&H_{\rm res}
    =   \sum_k \epsilon^{\rm L}_k c^{\dg}_kc_k
       + \sum_q \epsilon^{\rm R}_q d^{\dg}_q d_q ,
\label{H1c} \\
 &&H'  = \sum_{k,q} (T_{qk}+\chi_{qk}|a\ra\la a|)
         c^{\dg}_k d_q  + \mb{H.c.}
\label{Hp0}
\eea
\end{subequations}
The two terms in $H_{\rm res}$ are for electrons in
the two reservoirs (electrodes) labelled by ``L'' and ``R'',
respectively. The interaction Hamiltonian $H'$ here
describes  the electron tunneling through QPC,
e.g., from state $|q\ra$ in the
R-reservoir to state $|k\ra$ in the L-reservoir,
with the tuneling coupling amplitude of $(T_{qk}+\chi_{qk}|a\ra\la a|)$
that depends explicitly on the qubit state.
It is right this qubit-state dependence of the
tunneling amplitude that makes it possible to draw out the qubit
state information from the transport current through QPC.
In the above microscopic Hamiltonian,
the detector is described in terms of second quantization to address
the many particle nature,
meanwhile for the measured system (qubit) single particle description
is adopted since there is only one extra electron in it.
Here we denote the qubit state by $|a\ra$ and $|b\ra$, corresponding to the
electron locating in the left and right dot. In this work
we shall also introduce the qubit eigenstates $|1\ra$ and $|2\ra$,
which are the superpositions of the dot-states $|a\ra$ and $|b\ra$.

\subsection{Reduced description for the measured qubit}
\label{thsec2B}

  Quantum measurement can be characterized by dephasing
and relaxation of the measured system.
In this subsection, we present a unified QME
description for the reduced dynamics
of the qubit which is subjected to the measurement of a QPC.
Details of the formal derivation and the adopted approximations
are referred to Ref.\ \onlinecite{Li02}.
Here we only outline the key procedure and the main
results with respect to the measurement model in concern.

   It is well known that in weak coupling regime one can derive the QME
by carrying out a second-order cummulant expansion
with respect to the system-environment interaction Hamiltonian.
In our case, we treat the qubit-state-dependent tunneling Hamiltonian $H'$ as
perturbation, since it fully contains
the coupling information between the qubit and detector.
In the interaction picture with respect to the QPC reservoir Hamiltonian
$H_{\rm res}$ of  \Eq{H1c},
the interaction Hamiltonian $H'$ of \Eq{Hp0} becomes time dependent and reads
(setting $\Delta_{kq}\equiv\epsilon^{\rm L}_k-\epsilon^{\rm R}_q$)
\bea\label{H2}
H'(t)= \sum_{q,k} (T_{qk}+\chi_{qk}|a\ra\la a|)
    c^{\dg}_k d_q e^{i\Delta_{kq}t} + \mb{H.c.}
\eea
Making connection with the formalism developed in Ref.\ \onlinecite{Li02},
we denote $W_{qk}=T_{qk}+\chi_{qk}|a\ra\la a|$ and
$f_{qk}^{\dg}(t) = c^{\dg}_kd_q e^{i\Delta_{kq}t}$, which
are operators in the qubit and the stochastic
bath reservoirs subspaces, respectively,
and recast \Eq{H2} as
$H'(t) =\sum_{q,k}[W_{qk}f_{qk}^{\dg}(t)+W_{qk}^{\dg}f_{qk}(t)]$.
With this form and starting from the Liouville equation,
the QME satisfied by the reduced density matrix can be derived
after tracing out the microscopic degree of freedom of the QPC reservoirs,
precisely following the procedures in Ref.\ \onlinecite{Li02}.

  For simplicity, we assume $T_{qk}\equiv T$
and $\chi_{qk}\equiv \chi$, i.e.,
the tunneling amplitudes are reservoir-state independent.
Accordingly, the interaction Hamiltonian simplifies to
$H'(t) = Wf^{\dg}(t)+W^{\dg}f(t)$, where
\be\label{H3a}
  W = T+\chi |a\ra\la a|, \ \ \
  f^{\dg}(t) = \sum_{k,q} c^{\dg}_kd_q e^{i\Delta_{kq}t} .
\ee
The measurement current fluctuation-induced
dephasing and relaxation effects on the qubit
are characterized by the interaction bath correlation functions,
which in relation to the QPC detector shown in Fig.\ 1
can be carried out explicitly as
\begin{subequations}\label{Cpmt}
\bea
  \ti{C}^{(+)}(t) \equiv \la f^{\dg}(t)f(0) \ra
   = \sum_{k,q} e^{i\Delta_{kq}t} N_{{\rm L}k} (1-N_{{\rm R}q}),  \\
\ti{C}^{(-)}(t) \equiv \la f(t)f^{\dg}(0)\ra
  = \sum_{k,q} e^{-i\Delta_{kq}t} (1-N_{{\rm L}k}) N_{{\rm R}q}.
\eea
\end{subequations}
Here, $\la \cdots \ra$ stands for the statistical
average over both the left and right electron reservoirs,
which are assumed to be in the local thermal equilibrium,
with the Fermi-Dirac functions being given by
$N_{{\rm L}k}=[e^{\beta(\epsilon^{\rm L}_{k}-\mu_{\rm L})}+1]^{-1}$
and $N_{{\rm R}q}=[e^{\beta(\epsilon^{\rm R}_{q}-\mu_{\rm R})}+1]^{-1}$,
respectively.
Here, $\beta=1/(k_BT)$ is the inverse temperature, and
$\mu_{\rm L}$ and $\mu_{\rm R}$ are the chemical potentials
that relate to the applied voltage across
the detector by $\mu_{\rm L}-\mu_{\rm R}=eV$.
The interaction bath spectrum  is then defined as
the Fourier transform of the reservoir-electron
correlation function \cite{Li02},
\bea\label{Cpm}
  C^{(\pm)}(\pm \omega)
   =  \int^{\infty}_{-\infty}dt\, \ti{C}^{(\pm)}(t)e^{\pm i\omega t} .
\eea
It satisfies the detailed-balance relation of
$C^{(+)}(\omega)/C^{(-)}(-\omega)=e^{\beta(\omega+eV)}$.

  With the above clarifications, the unified QME
developed in Ref.\ \onlinecite{Li02} applied here to
describe the dephasing and relaxation of the
measured qubit can now be completely identified.
\bea \label{ME}
\dot{\rho} = -i {\cal L}\rho - {\cal R}\rho ,
\eea
with ${\cal L}(\cdots) \equiv [H_{\rm{qu}},(\cdots)]$ being
the quibit Liouvillian and
${\cal R}$ being the dissipation superoperator defined via
the following compact form \cite{Li02},
\bea\label{Rrho}
{\cal R}\rho = \frac{1}{2}\left[ W^{\dg},
      \ti{W}^{(-)}\rho-\rho\ti{W}^{(+)} \right] + \mb{H.c.}
\eea
Formally, $\ti{W}^{(\pm)}$ is related to the coupling operator $W$ as
\bea\label{Wpm}
\ti{W}^{(\pm)} = C^{(\pm)}(\pm{\cal L})W .
\eea
Here, $C^{(\pm)}(\pm{\cal L})$ is a superoperator, specified by
the qubit Liouvillian ${\cal L}$ and the interaction bath
spectrum $C^{(\pm)}(\pm \omega)$.
To obtain the explicit expressions of $C^{(\pm)}(\pm{\cal L})$,
we further adopt the continuum and
wide-band approximations for the QPC reservoir electrons.
Accordingly, the discrete summations in \Eq{Cpmt} can be
replaced by the continuous integrations,
$\sum_{k}\sum_q\rightarrow g_{\rm L} g_{\rm R}
\int\!\!\int d\epsilon^{\rm L}_{k}d\epsilon^{\rm R}_{q}$,
where the energy-independent density of states (DOS),
$g_{\rm L}$ and $g_{\rm R}$, are introduced
for the two reservoirs.
The analytical expressions for $C^{(\pm)}(\pm{\cal L})$
can then be readily integrated out as
\be \label{CpmL}
   C^{(\pm)}(\pm{\cal L}) = 2\pi g_{\rm L}g_{\rm R}
    \left[\frac{x}{1-e^{-\beta x}}
     \right]_{x=\pm({\cal L}+eV)} .
\ee
Equations (\ref{ME})--(\ref{CpmL}) constitute the QME formulation
that contains the full effects of measurement on the qubit
and will serve as the starting point of the following studies.

\subsection{Comments and discussions}
\label{thsec2C}

  Let us start with the high measurement voltage limit ($eV\gg {\cal L}$),
in which the applied measurement voltage
is much larger than the internal energy scale of the qubit.
In this case, the superoperator
$C^{(\pm)}(\pm {\cal L})$ of \Eq{CpmL}
reduces to a $c$-number,
\be \label{CL2C0}
 C^{(\pm)}(\pm {\cal L})\rightarrow
   C^{(\pm)}(0) = \pm 2\pi g_{\rm L}g_{\rm R} \frac{eV}{1-e^{\mp \beta eV}}.
\ee
With this approximation, \Eq{Rrho} recovers the QME
derived in Ref.\ \onlinecite{Goa01}; i.e.,
\be \label{A_QME}
  {\cal R}\rho \simeq -C^{(-)}(0){\cal D}[W]\rho
                    - C^{(+)}(0){\cal D}[W^{\dg}]\rho  ,
\ee
with ${\cal D}[W]\rho = W\rho W^{\dg} -\frac{1}{2} [W^{\dg}W,\rho]_+$
and ${\cal D}[W^{\dg}]\rho$ being defined similarly by swapping
between $W$ and $W^{\dg}$.
It is easy to show \cite{Goa01} that
the QME of \Eq{A_QME} is in fact also equivalent to the Bloch
equation derived by Gurvitz \cite{Gur97}.

  Our QME formulation in \Eqs{ME}--(\ref{CpmL})
is valid for arbitrary measurement voltage.
In contrast with the $c$-number of $C^{(\pm)}(0)$,
the operator nature of $C^{(\pm)}(\pm {\cal L})$ in
our QME formulation properly describes not only the dephasing
but also the important {\it energy exchange}
between the qubit and detector.
Physically, the quantum-measurement-induced dephasing and relaxation
on the qubit are originated from the
current fluctuations in the detector \cite{Sto02,Shn02,Sch02}.
As mentioned earlier, the current fluctuations are characterized
by the correlation functions $\ti{C}^{(\pm)}(t)$
of \Eq{Cpmt}, and their spectra satisfy the detailed-balance relation
of $C^{(+)}(\omega)/C^{(-)}(-\omega) = e^{\beta(\omega+eV)}$
at arbitrary temperature and measurement voltage [\cf \Eq{CpmL}].
Consequently, $C^{(\pm)}(\pm{\cal L})$ manifests
the backaction of the detector on the qubit by
correlating the measurement current fluctuations with
the qubit dissipations,
leading  to our QME in \Eqs{ME}--(\ref{CpmL})
satisfying the detailed-balance relation.
More specifically, the energy exchange
between the qubit and detector can be described as follows.
$C^{(+)}({\cal L})$ accounts for the
current fluctuations associated with electron tunneling in the detector
from the left to the right reservoirs,
accompanied by energy absorption from the qubit, while
$C^{(-)}(-{\cal L})$ corresponds to tunneling from the right
reservoir to the left one, accompanied by energy emission to the qubit.
The energy exchange characterized by
$C^{(\pm)}(\pm{\cal L})$, i.e.,
the current-fluctuation spectrum
$C^{(\pm)}(\pm\omega)$ correlated with qubit Liouvillian $\omega={\cal L}$,
is an essential feature hold by our QME that
manifests the important detailed-balance  relation.
This issue is further elaborated in the Appendix  by carrying out
the explicit expression of the QME in terms
of the qubit atomic operators.

\section{Measurement-induced dephasing and relaxation }
\label{thsec3}

   We are now in the position to highlight the essential role
and impact of the detailed balance on
the qubit dephasing and relaxation under measurement.
In order to have a close comparison with
the previous results that neglect the detailed balance
effects \cite{Gur97,Goa01},
instead of the QME given explictly
in the Appendix  in terms of the qubit atomic
operators $\sigma_z$ and $\sigma^{\pm}$,
in this section we would like to elaborate it
in the individual dot-state basis.
For clarity, our results will be presented
for the symmetric and asymmetric qubit cases,
separately, in the following two subsections.

\subsection{Symmetric case: $\epsilon_a=\epsilon_b$ }

In the individual dot-state representation $\{|a\ra,|b\ra \}$,
the coupling operator $W$ takes a matrix form
\be
W = \left[ \begin{array}{cc}
  T+\chi &  0   \\
  0      &  T   \\
\end{array}
\right] ,
\ee
and its spectral conjugate in \Eq{Wpm} is
\be
\ti{W}^{(\pm)} = \left[ \begin{array}{cc}
 \ti{W}^{(\pm)}_{aa}  &  \ti{W}^{(\pm)}_{ab}  \\
 \ti{W}^{(\pm)}_{ba}  &  \ti{W}^{(\pm)}_{bb}   \\
\end{array}
\right] .
\ee
The involving $\ti{W}^{(\pm)}$-matrix elements can
be evaluated readily via the standard operator algebra. To
simplify the notation, let us denote
\begin{subequations} \label{lambpm}
\bea
  \lambda_{\pm} &\equiv&
    [C^{(\pm)}(\pm\Delta) +C^{(\pm)}(\mp\Delta)]/4,
\\
   \bar{\lambda}_{\pm} &\equiv&
    [C^{(\pm)}(\pm\Delta) -C^{(\pm)}(\mp\Delta)]/4.
\eea
\end{subequations}
Here $\Delta=E_1-E_2$, with $E_1$ and $E_2$ being the qubit eigen-energies.

  For the symmetric qubit ($\epsilon_a=\epsilon_b=\epsilon_0$),
we have then
\begin{subequations}\label{Wpmab-s}
\bea
 \ti{W}^{(\pm)}_{aa}
&=& ( T+\chi/2) C^{(\pm)}(0)
   + \chi\lambda_{\pm} ,
\\
  \ti{W}^{(\pm)}_{bb}
&=& (T+\chi/2)C^{(\pm)}(0)   - \chi\lambda_{\pm} ,
\\
  \ti{W}^{(\pm)}_{ba} &=&  -\ti{W}^{(\pm)}_{ab} = \chi\bar\lambda_{\pm} .
\eea
\end{subequations}
Substituting \Eq{Wpmab-s} into the formal QME [\Eqs{ME} and (\ref{Rrho})],
the Bloch equations for the reduced density matrix in the dot-state
representation can be readily obtained.
For instance, the off-diagonal density matrix element satisfies
\bea\label{MEab1}
\dot{\rho}_{ab}
&=& -i(\epsilon_a-\epsilon_b)\rho_{ab}+i\Omega(\rho_{aa}-\rho_{bb})
\nl && -\chi^2 (\lambda_{+} + \lambda_{-})\rho_{ab}
\nl && -\frac{\chi^2}{2}(\bar\lambda_{+} - \bar\lambda_{-})
        (\rho_{aa}+\rho_{bb}).
\eea
For simplicity, we have assumed here the tunneling coefficients
$T$ and $\chi$ be real.
It is easy to see that in the absence of detailed balance, i.e.,
$C^{(\pm)}(\pm\Delta)\rightarrow C^{(\pm)}(0)$,
\Eq{MEab1} reduces to
\bea \label{MEab2}
\dot{\rho}_{ab} &=&
  -i(\epsilon_a-\epsilon_b)\rho_{ab}+i\Omega(\rho_{aa}-\rho_{bb})
\nl & &
  -\frac{\chi^2}{2}[ C^{(+)}(0)+C^{(-)}(0)]
  \rho_{ab},
\eea
which is nothing but the result derived in
Refs.\ \onlinecite{Gur97} and \onlinecite{Goa01}.

\begin{figure}\label{Fig2}
\begin{center}
\centerline{\includegraphics [scale=0.4] {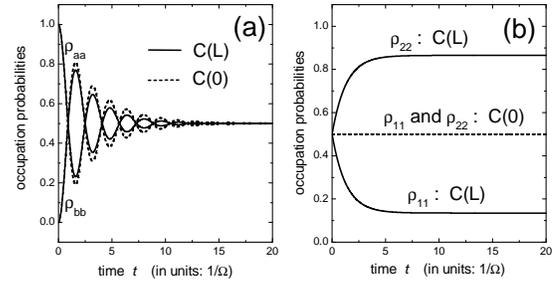}} \caption{
Measurement induced qubit relaxation in (a) the individual
dot-state representation, and (b) the eigenstate representation.
The results in the presence and the absence of the detailed
balance are symbolized by ``$C(\cal{L})$'' and ``$C(0)$'',
respectively. }
\end{center}
\end{figure}


  Under the quantum measurement, a pure state of the qubit
state evolves into a statistical mixture.
Figure 2 shows such evolution by plotting
the time-dependent occupation probabilities
on the individual dot states.
In the following numerical studies,
the relevant parameters are adopted as follows:
the applied voltage over the QPC $eV=\Omega$,
the inverse temperature $\beta=1/\Omega$,
the DOS in both electron reservoirs $g_{\rm L}=g_{\rm R}=2/\Omega$,
and the tunneling amplitudes $T=\Omega$ and $\chi=0.15\Omega$.
In the dot-representation as shown in
Fig.\ 2(a), despite certain quantitative difference in
short time scale, common final occupation probability
of $1/2$ in each dot is approached,
irrelevant to the detailed balance being satisfied or not.
Physically, due to the measurement-induced dephasing,
a transition occurs for the qubit electron tunneling
from the coherent to incoherent regime.
In the coherent regime, the tunneling
results in the well-known Rabi oscillations.
In the incoherent regime, no phase correlation
exists between the tunneling events,
and the readout appears telegraphic signals.
In the symmetric case,
owing to $\epsilon_a=\epsilon_b$, the final equal occupation
probability of $1/2$ in each dot is anticipated.
However, as shown by Gurvitz \cite{Gur97},
in the asymmetric case (i.e. for non-identical coupled dots),
final equal occupation probability of $1/2$ in each
dot would also be approached.
Similar confusing feature also existed
in the breakdown of the Anderson
localization, where equal occupation probabilities
on each site of the disordered chain were found \cite{Gur00}.
This peculiar feature is only valid in the limit of
large measurement voltage,
which causes an effective thermalization in terms of
an effective temperature.
The equal stationary occupation in a general asymmetric cases
is however unphysical; it violates
the detailed balance since it does not properly account
for the energy exchange between the
measured system and the detector.

  To reveal the significant implication of the detailed balance,
let us transform the result in Fig.\ 2(a)
into the qubit eigenstate representation, as shown in Fig.\ 2(b).
Viewing that initially the electron locates
in the left dot, which is equivalent to $1/2$
probability in each eigenstate of the symmetric qubit,
the constant dashed line in Fig.\ 2(b) indicates that
equal occupation probabilities on the two
eigenstates would be unaffected if
the detailed balance could be neglected.
However, the proper relaxation between the
eigenstates will result in quite different
occupation probabilities as shown
by the solid curves in Fig.\ 2(b).

\begin{figure}\label{Fig3}
\begin{center}
\centerline{\includegraphics [scale=0.4] {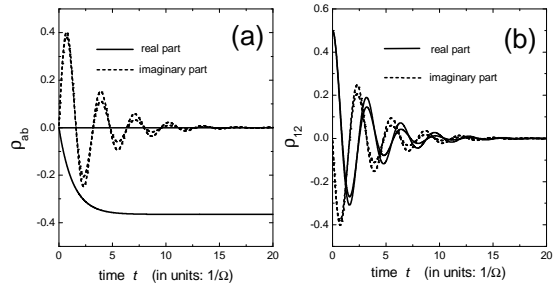}} \caption{
Measurement induced dephasing in (a) the individual dot-state
representation, and (b) the eigenstate representation. The real
part and imaginary part of the off-diagonal matrix element are
plotted by the solid and dashed curves, respectively. In (a), the
result in the absence of detailed balance is shown by the constant
zero solid line and the dashed curve with larger oscillation
amplitude. The other two curves are for the detailed-balance
preserved result. Similarly, in (b), the curves with smaller and
larger amplitudes correspond, respectively, to the
detailed-balance hold and un-hold results. }
\end{center}
\end{figure}

  Despite the drastic consequence of the detailed balance on
relaxation, our QME also stimulates an interesting issue in
dephasing. In Fig.\ 3(a) the dephasing behavior is described by
the off-diagonal density-matrix element in the dot-state
representation. We see that in the absence of detailed balance,
{\it complete dephasing} between the dot states takes place at
the long measurement time limit. However, in the presence
of detailed balance, the real part of $\rho_{ab}$ approaches a
nonzero constant.
We notice that a similar feature of nonzero off-diagonal
matrix element in dot-state basis appears also in
Ref.\ \onlinecite{Gur03} [see Eq.\ (11) there]
by coupling the qubit with an additional thermal bath.
In contrast, our result steams merely
from the coupling with the detector,
owing to the fact that our theory properly accounts for the energy
exchange between the qubit and detector,
and thus its consequence on dephasing and relaxation.
Mathematically, noting that $\rho_{aa}=\rho_{bb}=0.5$
as $t\rightarrow \infty$, the off-diagonal matrix element
approaches asymptotically to a non-zero value of
$\rho_{ab}(t) \rightarrow 0.5(\bar{\lambda}_{-}
  -\bar{\lambda}_{+})/(\lambda_{-} + \lambda_{+})$ via \Eq{MEab1},
rather than $\rho_{ab}(t)\rightarrow 0$ via \Eq{MEab2}.
Physically, our result based on the detailed-balance-preserved QME
indicates that the dot-state basis is not a proper representation
to show dephasing.
Under the weak measurement considered here, the qubit is weakly perturbed
by the detector and its eigenstates remain a good representation
to describe its dissipative dynamics.
In this qubit-Hamiltonian dominated regime,
complete dephasing is anticipated to take place
between the qubit eigenstates rather than the dot-states \cite{Sch98,note_1}.
In Fig.\ 3(b) we transform the off-diagonal density matrix element
$\rho_{ab}$ into $\rho_{12}$, i.e., from the dot-state basis to the
eigenstate basis.
A complete dephasing is observed satisfactorily
between the qubit eigenstates.

So far, we have restricted our discussion in the symmetric qubit,
and have already got insight in the impact of detailed balance on
the qubit relaxation and dephasing.
Below, we briefly show results for the
asymmetric qubit, where more apparent effects
can be observed.

\subsection{Asymmetric case: $\epsilon_a\neq\epsilon_b$}

In an asymmetric case, simple diagonalization of the qubit Hamiltonian
gives rise to the eigenstates
$|1\ra=\cos\frac{\theta}{2}|a\ra + \sin\frac{\theta}{2}|b\ra$ and
$|2\ra=\sin\frac{\theta}{2}|a\ra - \cos\frac{\theta}{2}|b\ra$;
see the Appendix.
We still denote the eigen-energy difference
by $\Delta=E_1-E_2$.
With the knowledge of eigenstates, one can straightforwardly evaluate
the operator $\ti{W}^{(\pm)}$ in the master equation.
In dot-representation, the result reads
\bea \label{Wpmab-as}
\ti{W}^{(\pm)}_{aa} &=& \left[T+\frac{\chi}{2}(1+\cos^2\theta) \right]
                        C^{(\pm)}(0)
        +\chi\lambda_{\pm}\sin^2\theta ,
\nl
\ti{W}^{(\pm)}_{bb} &=& \left( T+\frac{\chi}{2}\sin^2\theta \right)
                        C^{(\pm)}(0)
        - \chi\lambda_{\pm}\sin^2\theta,
\nl
\ti{W}^{(\pm)}_{ab} &=& \frac{\chi}{4}C^{(\pm)}(0)\sin2\theta
   - \chi(\bar{\lambda}_{\pm} + \lambda_{\pm}\cos\theta)\sin\theta,
\nl
\ti{W}^{(\pm)}_{ba} &=& \frac{\chi}{4}C^{(\pm)}(0)\sin2\theta
   + \chi(\bar{\lambda}_{\pm} - \lambda_{\pm}\cos\theta)\sin\theta.
\nl
\eea
In Fig.\ 4(a) and Fig.\ 4(b), the measurement-induced qubit state relaxation
is shown in (a) dot-state, and (b) eigenstate representations.
In the absence of detailed balance, we see in both representations the
qubit state relaxes to a statistical mixture with
equal probabilities on the two states of the qubit.
This peculiar  feature in asymmetric qubit
is owing to the equal probabilities
with which the transitions
from $|1\ra$ to $|2\ra$, and from $|2\ra$ to $|1\ra$,
take place in the absence of detailed balance, as shown by
the dominant Lindblad relaxation terms in the master equation in the
Appendix.
However, if the detailed balance is properly accounted for,
remarkably different statistical mixture will be approached
after the measurement, see the solid curves
in Fig.\ 4(a) and (b).
We anticipate that the relaxation behavior in Fig.\ 4(a)
can be demonstrated by future experiment.
For asymmetric qubit, the dephasing characteristics shown in
Fig.\ 4(c) and (d)
are similar to that in the symmetric qubit.
Again, complete dephasing takes place
between the eigenstates rather than the dot-states of the qubit
under weak measurement.

  It is desirable to compare the
measurement-induced
relaxation described in this work (\cf Figs.\ 2 and 4)
with that originated from coupling with an additional
thermal bath as discussed in Ref.\ \onlinecite{Gur03}.
In both cases, relaxations are resulted from energy exchange between
the system of interest and the environment.
In this sense, the relaxation induced here by a
measurement device should be similar to that by
a thermal bath.
This analogy is also discussed in Ref.\ \onlinecite{Moz02},
where the measurement voltage across the detector
is shown to be equivalent to
an effective temperature of thermal bath in certain sense.
Since the measurement in Ref.\ \onlinecite{Gur03}
is described as the previous work \cite{Gur97} which causes
only decoherence,
an additional bath is introduced there for relaxation.
It is expected that our QME approach allows
the measurement itself to generate the relaxation and
its consequences discussed in Ref.\ \onlinecite{Gur03}.

\begin{figure}\label{Fig4}
\begin{center}
\centerline{\includegraphics [scale=0.5] {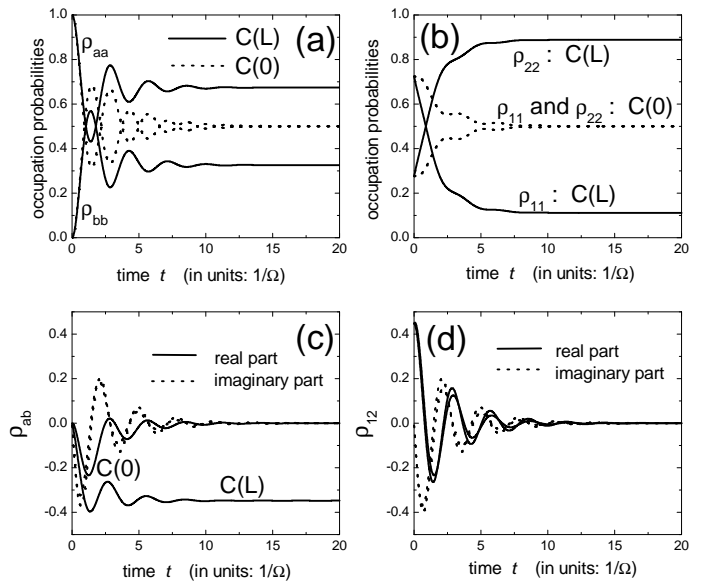}} \caption{
Measurement induced relaxation
and dephasing of an asymmetric qubit
(with dot-level offset $\epsilon_a-\epsilon_b=0.5\Omega$) in, respectively,
the individual dot-state representation [(a) and (c)],
and the eigenstate representation [(b) and (d)].
In (a)-(c), the results in the presence and absence
of the detailed balance
are symbolized by ``$C(\cal{L})$'' and ``$C(0)$''.
The qualitative feature of dephasing in (c) and
(d) is similar to the symmetric qubit,
and the corresponding figure description is referred to Fig.\ 3. }
\end{center}
\end{figure}

\section{Relaxation and dephasing rates}

In this section we carry out the analytical expressions
for relaxation and dephasing rates and discuss their characteristics
which depend on the measurement conditions.
Since the weak measurement under study is in the qubit-Hamiltonian
dominated regime, we present our analysis
in the qubit eigenstate representation (see the Appendix),
in which \Eq{ME} can be expressed as \cite{Li02}
\bea
\dot{\rho}_{jk} = -i\omega_{jk}\rho_{jk}
                  -\sum^2_{j',k'=1}{\cal R}_{jk,j'k'}\rho_{j'k'} ,
\eea
where $\omega_{jk}=E_j-E_k$, and the dissipation tensor elements reads
\be
{\cal R}_{jk,j'k'}= \left({\cal K}_{jk,j'k'}
                 +{\cal K}^*_{kj,k'j'}\right)/2,
\ee
with
\bea
  {\cal K}_{jk,j'k'}&=&
    \delta_{kk'}[W^{\dg}\ti{W}^{(-)}+W\ti{W}^{(+)\dg}]_{jj'}
\nl & &
   -[W^*_{kk'}\ti{W}_{jj'}^{(-)}+W_{k'k}\ti{W}_{j'j}^{(+)*}] .
\eea
These tensor elements have clear physical meaning.
For example, $-{\cal R}_{jj,kk}$ (with $j\ne k$)
amounts to transfer of the occupation
probability from $|k\ra$ to $|j\ra$,
while ${\cal R}_{jk,jk}$ describes the
dephasing between $|j\ra$ and $|k\ra$.
This can be further elucidated by making the
so-called secular approximation
which retains only the diagonal relaxation tensor elements,
such as ${\cal R}_{jj,kk}$ and ${\cal R}_{jk,jk}$.
In the qubit eigenstate basis $\{|1\ra,|2\ra\}$,
the secular approximation leads to the following Bloch equation
\begin{subequations} \label{rhoij}
\bea
  \dot{\rho}_{11}
&=& -\dot{\rho}_{22} = -\Gamma_1\rho_{11}+\Gamma_2\rho_{22},
\label{rhoijA} \\
  \dot{\rho}_{12} &=& -i\Delta\rho_{12} - \gamma_{12}\rho_{12} ,
\label{rhoijB}
\eea
\end{subequations}
where
\bea\label{Gamma12}
\Gamma_1 &\equiv& -{\cal R}_{22,11}=|W_{12}|^2
             [C^{(-)}(\Delta)+C^{(+)}(\Delta)] , \nl
\Gamma_2 &\equiv& -{\cal R}_{11,22}=|W_{21}|^2
             [C^{(-)}(-\Delta)+C^{(+)}(-\Delta)] , \nl
\gamma_{12}&\equiv& {\cal R}_{12,12}
        = \frac{1}{2} \left(\Gamma_1+\Gamma_2\right)
        + \frac{1}{2} \left(W_{11}-W_{22}\right)^2                 \nl
       & & \qquad\qquad \times \left[ C^{(-)}(0)+C^{(+)}(0)\right] .
\eea
The relaxation between $|1\ra$ and $|2\ra$ is
characterized by the evolution of
$\rho_z(t)\equiv \rho_{11}(t)-\rho_{22}(t)$.
From \Eq{rhoijA} it is easy to show that
\bea
\dot{\rho}_z(t)=-(\Gamma_1+\Gamma_2)[\rho_z(t)-\rho_z(\infty)],
\eea
which results in the solution
\bea
\rho_z(t)=\rho_z(\infty)+\left[ \rho_z(0)-\rho_z(\infty)\right]
          e^{-(\Gamma_1+\Gamma_2)t} .
\eea
Accordingly, the $T_1$-relaxation rate is obtained as
\bea \label{T1-1}
\frac{1}{T_1}
\! &=&\! \Gamma_1+\Gamma_2
\nl &=&\!
   \frac{g_{\rm L}g_{\rm R}}{2/\pi}
     [ F(eV+\Delta)+F(eV-\Delta) ]\chi^2\sin^2\theta ,
\eea
with
\be \label{Fx}
  F(x)\equiv x\coth(\beta x/2).
\ee
Similarly, the $T_2$-dephasing rate can be obtained as [\cf \Eq{rhoijB}]
\bea \label{T2-1}
\frac{1}{T_2}=\gamma_{12}=\frac{1}{2T_1}
    +\pi g_{\rm L}g_{\rm R} F(eV) \chi^2\cos^2\theta .
\eea
In this result, the first term stems from the relaxation-induced dephasing
and the second term describes the pure dephasing.
This identification can be simply understood as follows.
In the eigenstate representation (see the Appendix),
the qubit and interaction Hamiltonians read, respectively,
$H_{\rm{qu}}=\frac{\Delta}{2}\sigma_z$, and
$H'=W X=[(T+\chi)I+\frac{\chi}{2}(\cos\theta\sigma_z+\sin\theta\sigma_x)]X$.
From a master equation based analysis \cite{Yan00}, one can easily prove that
the $\sigma_x$-coupling
would cause the $T_1$-relaxation with rate $1/T_1$ given by \Eq{T1-1},
and simultaneously induce dephasing with rate $1/(2T_1)$.
Meanwhile, the $\sigma_z$-coupling
only results in pure dephasing with
rate given by the second term of \Eq{T2-1}.

\begin{figure}\label{Fig5}
\begin{center}
\centerline{\includegraphics [scale=0.4] {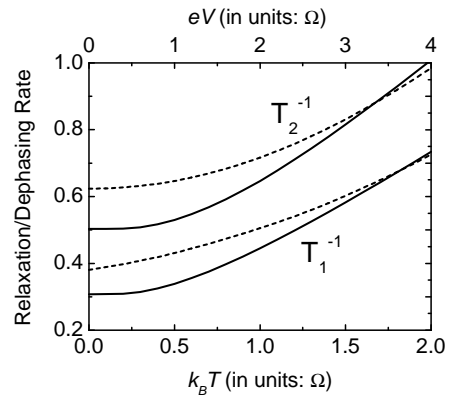}} \caption{
The measurement induced qubit
relaxation rate ($T_1^{-1}$) and
dephasing rate ($T_2^{-1}$) as functions of the temperature
for fixed voltage $eV=\Omega$ (solid curves),
and of the measurement voltage for fixed
temperature $k_BT=\Omega$ (dashed curves). }
\end{center}
\end{figure}

Equations (\ref{T1-1}) and (\ref{T2-1})
describe the dependence of the qubit relaxation
and dephasing rates on the various measurement parameters.
Most apparently, the rates depend on the
visibility parameter $\chi$ via $\propto\chi^2$,
which is the result in weak coupling regime, but
implies also that design of an appropriately
large $\chi$ is essential in order to
perform efficient measurement.
The dependence of the relaxation and dephasing rates on the
measurement voltage and temperature is numerically plotted in Fig.\ 5.
In general, both the applied voltage and temperature will enhance
the qubit relaxation and dephasing.
Dephasing will, in principle, benefit quantum measurement.
However, in practice the detector should be kept at very low temperatures,
since there exists a tradeoff between the signal and noise
strengths in the detector,
analysis on which has appeared in recent publications \cite{But02,Ave00}.
In what follows, based on \Eqs{T1-1} and (\ref{T2-1}),
we detail the voltage and temperature
dependence of the relaxation and dephasing rates under certain limits.
First, in the limit of zero bias voltage across the
detector, the temperature dependence
of the relaxation rate is characterized by
$T_1^{-1} \propto \coth(\Delta/2k_{B}T)$,
whereas the $\sigma_z$-coupling induced
pure-dephasing rate, i.e., the second term in \Eq{T2-1},
is $\propto k_B T$.
This difference in temperature dependence is due to that energy exchange
between the qubit and detector takes place during relaxation,
but there exists
no such exchange during pure dephasing.
Under zero bias voltage, the detector no longer plays role of
measurement, the qubit relaxation and dephasing
are merely caused by the zero-voltage
quantum and thermal fluctuations due
to random tunneling of electrons through the QPC.
Second, at zero temperature limit, the pure dephasing
rate linearly depends on the
measurement voltage by noting that $F(eV)\propto eV$.
Interestingly, the relaxation rate reduces to
$T_1^{-1}\propto (eV+\Delta+|eV-\Delta|)$, which linearly
depends on the voltage if $eV>\Delta$, but becomes a voltage-independent
constant when $eV<\Delta$.

\section{Conclusion}

In summary, we have studied the relaxation and dephasing of
a solid-state charge qubit
under quantum measurement of a mesoscopic detector.
Our treatment emphasizes in particular the energy
exchange between the qubit and detector
during the measurement process.
The measurement current fluctuation
is shown to have significant impact on not only the decoherence
but also the detailed-balance-preserved relaxation of qubit.
We have carried out both numerical and
analytical results for the qubit relaxation and dephasing,
and highlighted the new features which might be
relevant to future experiments.
Our unified QME approach is expected to be
generalized to a conditional version
which enables to study the readout statistics, and
to be unravelled by stochastic
wavefunction which can describe an individual
continuous measurement of a single qubit.
The work along these two lines is in progress and
will be published elsewhere.

\appendix*
\section{Elaboration on the detailed balance}

In this appendix we carry out the explicit operator form
for the relaxation super-operator
$R\rho$, from which the detailed balance retained by our
QME can be revealed clearly.
To this end, we express the qubit
operators in the eigenstate representation.
In general, for asymmetric qubit, the individual dot level
$\epsilon_a\neq\epsilon_b$,
we introduce $\epsilon=(\epsilon_a-\epsilon_b)/2$ for the dot-level offset,
and $\Delta=E_1-E_2$ for the qubit eigen-energy difference.
By a simple diagonalization of the qubit Hamiltonian,
the eigen-energies are obtained as $E_1=\sqrt{\epsilon^2+\Omega^2}$,
and $E_2=-\sqrt{\epsilon^2+\Omega^2}$.
Correspondingly, the eigenstates are
$|1\ra=\cos\frac{\theta}{2}|a\ra+\sin\frac{\theta}{2}|b\ra$,
and $|2\ra=\sin\frac{\theta}{2}|a\ra-\cos\frac{\theta}{2}|b\ra$,
where $\theta$ is introduced by
$\cos\theta=\epsilon/\sqrt{\epsilon^2+\Omega^2}$,
and $\sin\theta=\Omega/\sqrt{\epsilon^2+\Omega^2}$.
In the eigenstate basis $\{|1\ra,|2\ra\}$, the qubit
Hamiltonian reads $H_{\rm{qu}}=\frac{\Delta}{2}\sigma_z$,
and the coupling between the qubit and detector is described by
$H'=W X=[(T+\chi)I+\frac{\chi}{2}(\cos\theta\sigma_z+\sin\theta\sigma_x)]X$.
Here $I$ is the $2\times 2$ unit matrix, $X$ stands for
the tunneling operator
of the QPC, and the Pauli operators
$\sigma_z=|1\ra\la 1|-|2\ra\la 2|$, and $\sigma_x=|1\ra\la 2|+|2\ra\la 1|$,
which map the two-state qubit to a spin 1/2 particle.

In terms of the Pauli matrices,
the formal QME, Eq.\ (5), can be recast to an explicit form with
\bea\label{Rrho_1}
{\cal R}\rho &=& \eta_1^2 C(0)[\sigma_z,[\sigma_z,\rho]]
        + \eta_2^2 [\sigma_x,\ti{Q}_x\rho-\rho\ti{Q}^{\dg}_x]
\nl & &
       + \eta_1\eta_2 C(0)[\sigma_x,[\sigma_z,\rho]]
\nl & &
       + \eta_1\eta_2 [\sigma_z,\ti{Q}_x\rho-\rho\ti{Q}^{\dg}_x] ,
\eea
where $\eta_1=\frac{\chi}{2}\cos\theta$, $\eta_2=\frac{\chi}{2}\sin\theta$,
$C(0)=C^{(+)}(0)+C^{(-)}(0)$, and
$\ti{Q}_x\equiv C(-{\cal L})\sigma_x$
with $ C(-{\cal L})\equiv C^{(+)}(-{\cal L})+C^{(-)}(-{\cal L})$.
In the right hand side of \Eq{Rrho_1}, the first term
describes the $\sigma_z$-coupling induced
pure dephasing, the second term dominantly contributes
the $T_1$-relaxation as well as its associated
dephasing owing to the $\sigma_x$-coupling,
and the last two terms stem from the correlation
between the two couplings which have minor contribution
to the dissipative dynamics.
Due to the dominant $T_1$-rate contribution of the second term,
we further express it into a Lindblad-type form,
\bea\label{relax_1}
& & [\sigma_x,\ti{Q}_x\rho-\rho\ti{Q}^{\dg}_x]
\nl &=&
   -2C(-\Delta){\cal D}[\sigma^+]\rho
  -2C(\Delta){\cal D}[\sigma^-]\rho
\nl &\ &
   -[C(-\Delta)+C(\Delta)]
    (\sigma^-\rho\sigma^-+\sigma^+\rho\sigma^+).
\eea
where the Lindblad super-operators are defined as
\begin{subequations} \label{Dsigpm}
\bea
{\cal D}[\sigma^+]\rho &=& \sigma^+\rho\sigma^- - \frac{1}{2}
           \{\sigma^-\sigma^+,\rho \},  \\
{\cal D}[\sigma^-]\rho &=& \sigma^-\rho\sigma^+ - \frac{1}{2}
           \{\sigma^+\sigma^-,\rho \} .
\eea
\end{subequations}
In deriving this result, we have carried out
$\ti{Q}_x=C(-\Delta)\sigma^++C(\Delta)\sigma^-$, where
$\sigma^{\pm}=\frac{1}{2}(\sigma_x\pm i\sigma_y)$.
In this context, simple algebras were used as follows.
Note that ${\cal L}\sigma_x\equiv [H_{\mb{qu}},\sigma_x]
=\frac{\Delta}{2}(2i)\sigma_y$,
${\cal L}^2\sigma_x=(\frac{\Delta}{2})^2(2i)(-2i)\sigma_x$, and so on.
It then follows that the action of an arbitrary function of the
Liouvillian operator ${\cal L}$,
say, $C({\cal L})$, on $\sigma_x$ reads
$C({\cal L})\sigma_x=C_1(\Delta)\sigma_x+iC_2(\Delta)\sigma_y$,
with $C_1(\Delta)=[C(\Delta)+C(-\Delta)]/2$, and $C_2(\Delta)
  =[C(\Delta)-C(-\Delta)]/2$.

The terms ``$\sigma^-\rho\sigma^-$'' and ``$\sigma^+\rho\sigma^+$''
in \Eq{relax_1} are out of the rotating-wave-approximation and
their effects are small comparing with the
Lindblad terms with ${\cal D}[\sigma^{\pm}]\rho$ [\Eq{Dsigpm}].
Physically, ${\cal D}[\sigma^-]\rho$ describes quantum jump
from the upper qubit state
$|1\ra$ to the lower state $|2\ra$, and ${\cal D}[\sigma^+]\rho$ vice versa.
With satisfaction, the corresponding
jump probability $C(\pm\Delta)$ precisely relates
the qubit jump to the electron tunneling in the detector in the presence of
energy-quanta (i.e., $\Delta$) emission (absorption).
Note that this energy exchange is essential to ensure the detailed balance.
Denoting the occupation probabilities on the qubit states $|1\ra$ and $|2\ra$
by $P_1$ and $P_2$, at the stationary mixture state, the dominant term of
Eq.\ (A2) leads to $P_1/P_2=C(-\Delta)/C(\Delta)$. This is nothing but
a generalization of the usual detailed balance relation for coupling with
a thermal bath. Here the measurement voltage plays
certain role of an effective
temperature. This result is also in complete
consistence with the rate analysis
in Sec.\ IV, see \Eqs{rhoij} and (\ref{Gamma12}).
If we neglect the energy exchange, say, let $C(\pm\Delta)\rightarrow C(0)$,
the detailed balance is broken down, and equal occupation probabilities
on the qubit states are inevitably resulted in as in
the previous literatures \cite{Gur97,Goa01}.

\begin{acknowledgments}
  Support from the Major State Basic Research Project
No.\ G001CB3095 of China, the Special Fund for``100 Person Project"
from Chinese Academy of Sciences,
and the Research Grants Council of the Hong Kong Government
is gratefully acknowledged.
\end{acknowledgments}





\begin{thebibliography}{99}
\bibitem{note_1}
In the regime of strong (projective) measurement,
the qubit-detector interaction
Hamiltonian will dominate the qubit dynamics. Accordingly,
complete dephasing between the individual dot states
(i.e.\ the eigenstates of the observable) is anticipated.
However, the usual perturbative-theory-based QME approach does not work
in this regime.
\end{thebibliography}
\end{document}